\documentclass[apjl]{emulateapj}
\slugcomment{{\sc Accepted to ApJ Letters:} 7 June 2012} 

\usepackage{graphicx}

\newcommand{\degree}{\ensuremath{^\circ}}

\shorttitle{A Likely Close-In Low-Mass Stellar Companion to the Transitional Disk Star HD 142527}
\shortauthors{Biller et al.}

\begin{document}


\title{A Likely Close-In Low-Mass Stellar Companion to the Transitional Disk Star HD 142527}


\author{Beth Biller\altaffilmark{1}, Sylvestre Lacour\altaffilmark{2}, Attila Juh\'asz\altaffilmark{3}, Myriam Benisty\altaffilmark{1},
  Gael Chauvin\altaffilmark{4,1},  Johan Olofsson\altaffilmark{1},  J\"org-Uwe Pott\altaffilmark{1}, Andr\'e M\"uller\altaffilmark{1}, Aurora
  Sicilia-Aguilar\altaffilmark{5}, Micka\"el Bonnefoy\altaffilmark{1}, Peter Tuthill\altaffilmark{6}, Philippe
  Thebault\altaffilmark{2}, Thomas Henning\altaffilmark{1}, Aurelien Crida\altaffilmark{7}}      

\altaffiltext{1}{Max-Planck-Institut f\"ur Astronomie, K\"onigstuhl 17, 69117 Heidelberg, Germany}
\altaffiltext{2}{LESIA, CNRS/UMR-8109, Observatoire de Paris, UPMC, Universit\'e Paris Diderot, 5 place Jules Janssen, 92195 Meudon, France}
\altaffiltext{3}{Leiden Observatory, Leiden University, P.O. Box 9513, 2300 RA Leiden, The Netherlands}
\altaffiltext{4}{UJF-Grenoble 1 / CNRS-INSU, Institut de Plan\'etologie et d’Astrophysique de Grenoble (IPAG) UMR 5274,
Grenoble, F-38041, France}
\altaffiltext{5}{Departamento de F\'{\i}sica Te\'{o}rica, Facultad de Ciencias, Universidad Aut\'{o}noma de Madrid, 28049 Cantoblanco, Madrid, Spain}
\altaffiltext{6}{School of Physics, University of Sydney, NSW 2006, Australia}
\altaffiltext{7}{Universit\'e de Nice - Sophia antipolis / C.N.R.S. / Observatoire de la C\^ote d'Azur, Laboratoire Lagrange (UMR 7293), Boulevard de l'Observatoire,
  B.P. 4229  06304 NICE cedex 04, France}

%\affil{MPIA, etc.}

%\author{C. D. Biemesderfer\altaffilmark{4,5}}
%\affil{National Optical Astronomy Observatories, Tucson, AZ 85719}
\email{biller@mpia.de}

%\and

%\author{R. J. Hanisch\altaffilmark{5}}
%\affil{Space Telescope Science Institute, Baltimore, MD 21218}


%\altaffiltext{1}{Visiting Astronomer, Cerro Tololo Inter-American Observatory.
%CTIO is operated by AURA, Inc.\ under contract to the National Science
%\altaffiltext{2}{Society of Fellows, Harvard University.}
%\altaffiltext{3}{present address: Center for Astrophysics,
 %   60 Garden Street, Cambridge, MA 02138}
%\altaffiltext{4}{Visiting Programmer, Space Telescope Science Institute}
%\altaffiltext{5}{Patron, Alonso's Bar and Grill}


\begin{abstract}
With the uniquely high contrast within 0.1''  ($\Delta$mag(L') = 5-6.5 
magnitudes) 
available using Sparse Aperture Masking (SAM) with NACO at VLT, 
we detected asymmetry in the flux from the Herbig Fe star 
HD 142527 with a barycenter emission situated at
a projected separation of  88$\pm$5 mas (12.8$\pm$1.5~AU at 145 pc)
and flux ratios in H, K, and L' of   
0.016$\pm$0.007, 0.012$\pm$0.008, 0.0086$\pm$0.0011 
respectively (3-$\sigma$ errors) relative to the primary 
star and disk.  After extensive closure-phase modeling, we
interpret this detection as a close-in, low-mass stellar companion with an estimated  
mass of $\sim$0.1-0.4 M$_{\odot}$.  HD 142527 has a complex disk
structure, with an inner gap imaged in both the near and mid-IR as well as a
spiral feature in the outer disk in the near-IR.
This newly detected low-mass stellar 
companion may provide a critical explanation of the observed disk structure.
\end{abstract}





\section{Introduction}

Transition disks may trace a key step in the process of forming planets
and dissipating primordial stellar disks. 
Transition disks are primordial disks 
characterized by weak mid-IR emission (at $\sim$15 $\mu$m) 
relative to the Taurus median
spectral energy distribution
\citep[i.e.~the~median~SED~of~primordial~disks~in~the~young~($<$2~Myr)
~Taurus~star-forming~region]{Naj07}.
%(Najita et al. 2007).
%Transition disks may trace a key step in the process of forming planets
%and dissipating primordial stellar disks.  
A number of transition disks
possess gaps either posited from SED studies
or directly imaged \citep[]{And11, Fuk06, Pot10, Bro09}
which may be due to clearing of dust by a forming planet or brown
dwarf and thus may produce the observed mid-IR deficit.
%Accretion onto these bodies will indeed open a gap in the density
%distribution, resulting in inner and outer disks.
Transition disks have therefore been popular targets for high-contrast 
high-resolution imaging planet searches.  

While transition disks have previously been searched for planets using 
high contrast imaging techniques (e.g. adaptive optics and
coronagraphy), most of these techniques currently do not extend to the 
inner 0.1'' which corresponds to the crucial planet-forming
regions ($\sim$10 AU) of the disk  
for objects at distances $>$100 pc.  Only interferometric techniques 
possess the resolution to reveal these inner regions \citep[see~e.g.][]{Pot10}
%(see e.g. Pott et al. 2010) %, who reached flux ratios of 0.05 at
%separations from 2.5-30 mas).  
The interferometric technique of Sparse-Aperture
Masking (SAM) uniquely allows us to both probe the inner 0.1'' of transitional 
disks and reach high contrasts of $\Delta$mag(L') =5-6.5 mag.  Probing this region enables us to
understand how planets form in their native disk and how they impact the
surrounding material.  Indeed, observations using SAM have already 
yielded two planetary or brown dwarf candidate companions
 to transition disk stars \citep[]{Hue11, Kra12}.

The transition disk star HD 142527 is a Herbig Fe star with
spectral type of F6 IIIe \citep[]{Hou78, Hen76, Wae96}.
HD 142527 notably possesses a very complex and interesting disk 
that has long been posited as a possible site of planet formation.
Recent SED modeling and VISIR imaging suggests a disk gap from 30 -
130 AU \citep[]{Ver11}.  The outer edge of the gap as well as a spiral feature in the outer disk 
have been imaged in the near-IR \citep[]{Fuk06}.
\citet[][]{Fuk06} also find an offset of 20 AU
between the star center and disk center, which they
posit is caused by an unseen eccentric binary companion.  
\citet[][]{Bai06} note this system as a possible
(but unconfirmed) binary detection from spectro-astrometry.
The disk of HD 142527 also possesses an extremely high fraction 
of crystalline silicates, possibly formed by a massive companion 
inducing spiral density waves in the disk material \citep[]{Boe04}.
Here we report the discovery of a likely close-in, low-mass stellar companion 
(12.8$\pm$1.5~AU at 145 pc and flux ratios in H, K, and L' of   
0.016$\pm$0.007, 0.012$\pm$0.008, and 0.0086$\pm$0.0011 
respectively) to this star.  This is the first confirmation of the 
binarity of HD 142527 and may provide a critical explanation of the observed disk structure.

\section{Stellar Parameters}

%Waelkens et al. (1996) classified HD 142527 as a Herbig Fe star.
Hipparcos measurements for HD 142527 yield a distance of 230$^{+70}_{-40}$ pc
\citep[]{vLe07}.
%, i.e. with quite significant uncertainty (van den Ancker et al. 1998).
Alternately, HD 142527 has been associated with both the 
Sco OB-2 association \citep[]{Ack04} and 
Upper Centaurus Lupus \citep[]{deZ99, Tei00}.
Membership in either association (indeed the two both are part
of the larger Sco-Cen association) place HD 142527 at a distance of 140-145 pc 
\citep[]{deZ99} and an age of 2-10 Myr.  
We consider the evidence of association within Sco-Cen to be very strong
 and thus adopt a distance of 145$\pm$15 pc, which still lies 
within 2-$\sigma$ of the rather uncertain Hipparcos measurement.

\citet[]{Ver11} obtain a stellar luminosity of 15$\pm$2 
L$_{\odot}$ from comparison to \citet[]{Kur91} photospheric 
models for the object spectral type of F6 III.
They correct the stellar luminosity to 20$\pm$2 L$_{\odot}$
after taking into account a model-dependent grey extinction component of the 
disk and halo and obtain updated stellar parameters by 
comparing the position of this star in the Hertzsprung-Russell
diagram to the PMS evolution tracks of \citet[]{Sie00}.
Here, we adopt the stellar parameters from \citet[]{Ver11} --
specifically, a stellar mass of 2.2$\pm$0.3 M$_{\odot}$
and an age of 5$^{+8}_{-3}$ Myr
consistent with membership in the Sco OB-2
association (Table~\ref{tab:photom}).

\section{Observations and Data Reduction}

HD 142527 was observed on 10 March 2012 with VLT
NACO\footnote[1]{program id: 088.C-0691(A)}.
Observations were taken in the H, K, and L' bands 
($\lambda_{\mathrm{L'}}$= 3.80 $\pm$ 0.31 $\mu$m) using the ``7 holes'' 
aperture mask and the IR wavefront sensor (WFS).  The observing
log is presented in Table~\ref{tab:obslog}.   The target was 
observed in each band for 30 min to 2 hrs.

The use of the ``7 holes'' \citep[C7-892,][]{Tut10}
aperture mask transforms the telescope into a 
Fizeau interferometer. The point spread function is a complex
superposition of fringes at given spatial frequencies. In specific cases,
pupil-masking can outperform more traditional differential imaging
for a number of reasons \citep[][]{Tut06, Lac11a}.
First, the masks are designed to have
nonredundant array configurations that permit phase deconvolution;
slowly moving optical aberrations not corrected by the AO
can be accurately calibrated. Second, the mask primarily rejects
baselines with low spatial frequency and passes proportionately
far more baselines with higher $\lambda$/B 
(where B is baseline length) resolution than does an orthodox
fully filled pupil. Third, high-fidelity recovery of phase
information allows “super resolution”, with a marginal loss of
dynamic range up to $\lambda$/2D (where 
D is the mirror diameter). The principal drawback is a loss
in throughput so that photon and detector noise can affect the
signal-to-noise ratio even where targets are reasonably bright for
the AO system.  The effective field-of-view of SAM
is determined by the shortest baseline so that the technique is
not competitive at separations that are greater than several
times the formal diffraction limit.   For SAM observations with 
the ``7 holes'' mask with VLT NACO, the field-of-view is
300 mas in H band, 400 mas in K band, and 600 mas in L band.
For more details on the SAM mode, please see e.g. \citet[]{Lac11b, Tut10}.

The HD 142527 observations were processed with both the 
Sydney FFT and Observatoire de Paris SAMP pipelines \citep[]{Tut00, Lac11b}.
Both the $\chi^2$ map and the phase in the UV
plane \citep[extracted~from~the~closure phases,~see][]{Hue11}
show an asymmetry typical of point sources (at the resolution of the telescope, see
left panel of Figure 1).  Therefore, we fitted the closure phases with a
model of two point-like objects (the star and a companion of lesser
flux).  In all three bands, the best fit model to the 
closure phases shows a point-like asymmetry at 88$\pm$5 mas from the 
central star with flux ratios in H, K, and L' of   
0.016$\pm$0.007, 0.012$\pm$0.008, 0.0086$\pm$0.0011 
respectively (3-$\sigma$ errors) relative to the primary star and
disk.   The $\chi^2$ map of the L' band data is shown in the
right panel of Fig. 1. The fit on several of the 35 closure phases
triangle is plotted in Fig. 2 (L' band data). 

\section{Results}

%We detect a likely close companion to HD 142527 with separation 
%of  88$\pm$5 mas (12.8$\pm$1.5~AU at 145 pc)
%and flux ratios in H, K, and L' of   
%0.016$\pm$0.007, 0.012$\pm$0.008, 0.0086$\pm$0.0011 
%respectively (3-$\sigma$ errors) relative to the primary 
%star and inner disk.

\subsection{Photometry}

We adopt the HKL magnitudes from \citet[]{Mal98},
since no L band data is available
from 2MASS.
%and almost all SED
%modeling attempts for this system have used these
%values.  
We note, however, that there is a significant divergence
between the reported 2MASS photometry and the 
\citet[]{Mal98} photometry.  The 2MASS photometry 
is brighter than the \citet[]{Mal98} photometry
by 0.1-0.3 mag, which may suggest variability for this system.
No errors are provided for the \citet[]{Mal98}
photometry; we assume error bars are similar 
to the 2MASS photometry.

Raw photometry for this system is comprised of light from 
three components -- primary star, 
secondary companion, and disk.  Some portion of the disk
(the ``outer disk'') lies outside of the SAM field of view.
From the SED model of \citet[]{Ver11} we estimate that
the outer disk comprises $\sim$10$\%$ or less of the total
system flux at HKL'.  Thus, we do not correct the raw 
photometry to remove the outer disk component.
Magnitudes in H, K, and L' for the star+disk are presented 
in Table~\ref{tab:photom}.  

The detected companion has flux ratios in H, K, and L' of   
0.016$\pm$0.007, 0.012$\pm$0.008, 0.0086$\pm$0.0011 
respectively (3-$\sigma$ errors).  To account for errors in the 
initial photometry as well as our measured SAM flux ratio,
we adopt a Monte Carlo approach.  We simulated an ensemble 
of 10$^{6}$ observations per band, with photometry and flux ratios drawn 
from Gaussian distributions centered on the measured 
values and with $\sigma$ drawn from the reported errors.
The apparent magnitude in each band is given as the 
median of this ensemble, with error bars drawn from the 
standard deviation of the same ensemble. 
Thus, the measured flux ratios and primary star photometry 
 correspond to apparent magnitudes of 
10.5$\pm$0.2, 10.0$\pm$0.3, 9.1$\pm$0.1
in HKL' \citep[Table~\ref{tab:photom},~in~CIT~bandpasses,][]{Eli82}.
%converted to the CIT bandpasses
%(Elias et al. 1982, 1983) using the transforms from the 2MASS all-sky data release in order 
%for easy comparison with the models of Baraffe et al. (1998)).
The companion appears anomalously bright
in L'.  While the H-K color are similar to what would be expected
for a young red companion, K-L' $\sim$0.9 mag, diverging significantly 
from the expected value of $\sim$0.4 mag \citep[]{Bar98}.
Companion fluxes in these bandpasses,
along with the full SED for the system are plotted in 
Fig.~\ref{fig:sed}.  

We employed a similar Monte Carlo approach 
in converting from apparent to absolute magnitudes.  For our 
ensemble of 10$^{6}$ simulated objects, we simulate 
corresponding distances drawn from a Gaussian
centered at 145 pc and with $\sigma$ of 15 pc.
Absolute magnitudes are also reported in Table~\ref{tab:photom}.

\subsection{Probability of Chance Alignment}

We estimated the likelihood that this companion is an unrelated
background or foreground object using source counts from the 2MASS survey. Within a 
1 degree radius of the primary, 2MASS detects 1918 objects with H of 10.7 mag or brighter 
and 1505 objects with Ks of 10.0 mag or brighter.   
Thus, adopting the approach of \citet[]{Bra00}, 
in particular, their equation 1, we estimate the probability of finding an unrelated source at
least as bright as the observed companion within 0.088\arcsec of the
primary to be $\sim$1.1$\times$10$^{-6}$ in H band and 
$\sim$8.3$\times$10$^{-7}$ in Ks.  We also considered simulated 
stellar populations along this line of sight 
(Galactic latitude and longitude of 335.6549\degree, +08.4804\degree) using the Besan\c{c}on
Galactic population synthesis models \citet[]{Rob03}.
This line of sight is directly into the Galactic bulge, so the 
models yield 882 background sources per square degree brighter than 
K=10.5 mag.  However, the chances of finding one of these within 0.088''
of the primary are still vanishingly small --
$\sim$1.6$\times$10$^{-6}$ -- and these objects are predominantly 
M giant stars, with 
considerably bluer expected colors 
\citep[for~a~M5III~star,~H-K~=~0.29~mag~and~K-L'~=~0.22 mag,][]{Tok00}
than measured for the
detected companion.  It is therefore extraordinarily unlikely
that the companion is unrelated to the primary, although proper motion
confirmation in a year will be necessary to finally determine
this.

\subsection{Mass Estimate}

Estimated masses for both system components are highly dependent on
adopted age.  Certainly the HD 142527 system is quite young, but 
whether it is 1 Myr or 10 Myr makes a critical difference in the mass 
estimate for the faint companion.  Here, we adopt a similar age range
as \citet[][]{Ver11}, which is dependent on membership in the Sco OB-2 
association.  

%HD 142527 has also been suggested as a possible
%member of the somewhat younger Upper Centaurus Lupus association 
%as well (de Zeeuw et al. 1999, Teixeira et al. 2000).  A younger age
%will produce a lower mass estimate for the faint component; by adopting 
%a slightly older age we are thus putting upper limits on the possible
%mass of the companion.

We again adopt Monte Carlo methods to account for the range of 
possible ages for this system.  An ensemble of 10$^{6}$ possible ages are drawn
from a Gaussian in log space, centered on log(age) = 6.7 and with 
$\sigma$(log(age)) = 0.4.  We then interpolate with age and single
band absolute magnitude to find the best mass for the companion 
from the models of \citet[]{Bar98}.  The resulting mass 
distributions are presented as histograms in Fig.~\ref{fig:masshist}.
It is apparent that the mass estimate for the companion is not 
well constrained at these ages but is most likely to lie in the range from
0.1-0.4 M$_{\odot}$.  We estimate a best mass estimate for each band 
of 0.28$\pm$0.15 M$_{\odot}$, 0.34$\pm$0.19 M$_{\odot}$, 
and 0.60$\pm$0.29 M$_{\odot}$ in HKL' respectively.
However, the mass distributions from our Monte Carlo simulations 
are highly non-Gaussian with significant probability to find 
considerably higher companion masses.
All mass estimates are within 2-$\sigma$ of each other,
but the L' band mass estimate is particularly high and 
we note that the companion appears anomalously bright
in L'.   While the H-K colors are similar to what would be expected
for a young red companion, K-L' $\sim$0.9 mag, considerably divergent 
from the expected value of 0.4 mag.  We thus do not attempt to estimate
spectral type using H-K and K-L' colors.
The models of \citet[]{Sie00} yield a similar mass range 
for the companion of 0.1-0.4 M$_{\odot}$ for ages of 2-12 Myr.
The models themselves are highly dependent on age; to illustrate this,
we plot isomass contours \citep[from~the~models~of][]{Bar98}
as a function of absolute H magnitude and age in Fig.~\ref{fig:masshist}. 

%We estimated the primary mass by first estimating the absolute magnitudes 
%for the bare photosphere of the star using the multi-component
%SED model from Verhoeff et al. (2011) and then plugging in these
%estimates to the models of Siess et al. (2000).
%According to its position on the color-magnitude diagram and using 
%our measured HKL' photometry and adopting V-R=0.03 mag, we find a range of 
%masses from 2.0 to 2.5 M$_{\odot}$ for ages commensurate to our estimated 
%age range.  We thus adopt the same mass range of 2.2$\pm$0.3 M$_{\odot}$
%for the primary as Verhoeff et al. (2011).

\subsection{Constraints on the Orbit}

We estimate the semimajor axis of HD 142527B's orbit from its
observed separation. Assuming a uniform eccentricity distribution
between 0 $<$ e $<$ 1 and random viewing angles, 
\citet[]{Dup10} compute a median correction factor between
projected separation and semimajor axis of 1.10$^{+0.91}
_{-0.36}$ (68.3\% confidence limits). Using this, we derive a semimajor axis of
14$^{+12}_{-5}$ AU.
% based on its observed separation on 10 March 2012.
While the mass estimate for the companion is quite uncertain,
we adopt the K band value as characteristic and adopt
a total system mass of 2.54$\pm$0.35 M$_{\odot}$.
Our derived semi-major axis estimate corresponds to 
an orbital period estimate of 33$^{+42}_{-18}$ years.

After a year, we expect up to 20 mas of orbital motion on the sky
for the companion, which is easily detectable with SAM.
The degree of motion observed will put important constraints 
on the mass of the companion (after adopting an estimate 
of the mass of the primary) and will provide a key datapoint 
for an eventual dynamical mass determination for this system.
Assuming a coverage of one third of an orbit is necessary for a good
orbital determination \citep[]{Dup10}, such a determination may be possible in 
10 year timescales for this system.  HD 142527A has a measured
proper motion of -11.19$\pm$0.93 mas in RA and
-24.46$\pm$0.79 in DEC -- i.e. about $\sim$20 mas on-sky motion 
in a year, with 3-$\sigma$ astrometric errors of $\sim$5~mas.
Thus, we will also be able to completely rule out
the extremely unlikely case that the companion is an unrelated
background object.

\section{Discussion}

The existence of an inner binary has been predicted for 
HD 142527 \citep[]{Fuk06, Bai06}
but this is the first confirmation of the binary companion.
The inner binary likely explains the 20 AU offset observed
by \citet[]{Fuk06} between the primary and the disk center
\citep[]{Pic08, Nel03}.

The discovery of the inner binary provides an important update for 
modeling efforts of the structure of the inner disk.  The modeling 
efforts of \citet[]{Ver11} find a flat, dusty inner disk 
from 0.3 to 30 AU (outer radius supported as well by 
marginally resolved VISIR imaging), an optically thin halo from 0.3 to
30 AU, and an outer disk starting at 130 AU.
The current projected separation for the companion, $\sim$13 AU,
place it right inside the modeled inner disk!  Thus, it is likely that the 
companion may produce a cleared ring within the inner disk.
%With a flux ratio of $\sim$100:1 between the primary and the 
%secondary and $\sim$1:1 between the primary and the disk, 
%the secondary does not contribute greatly to the total SED 
%of the system, but it may contribute significantly to the structure
%of the inner disk.  
Models of the physical properties of disks 
around eccentric stellar binaries often show entirely cleared inner
disks \citep[]{Pic08}, which may not be the case here.

%Updated SED studies which take into account the 
%inner binary will be necessary to estimate inner disk properties.

The HD 142527 disk is notable for being 
comparably bright or even brighter
than the primary star at infrared wavelengths.  At H and K bands, the
star is as bright or brighter than the disk, whereas at L band, 
the disk is considerably brighter than the star.   We consider 
all components of the \citet[]{Ver11} SED model in 
deriving photometry for the companion.
However, if a portion of the inner disk does not contribute to the
brightness of the central source, this may be the reason that our L' band 
flux measurement for the companion is anomalously bright. 
Alternatively, the very bright measured L' band magnitude may suggest that the secondary has 
its own small circumstellar disk (with possible 
accretion onto the secondary) or that the relatively massive secondary 
may produce local disk heating.  Transient heating caused 
by the secondary inducing spiral shock waves in the disk material 
could also be the source of the high fraction of crystalline
sillicates found at large radii in this disk \citep[]{Boe04}.

The large cavity observed in the HD 142527 disk may be the signature 
of an unseen companion interacting with the inner binary.  
A binary + planet system can open a much larger  
gap than can be formed by the binary by 
itself \citep[][]{Nel03, Kle08a, Kle08b, Kle12}.  
\citet[]{Nel03} note that in the case of an eccentric inner
binary system, the circumbinary disk itself can become 
eccentric, ending the inner migration of the planet and 
producing a stable orbital configuration.  This seems a likely
explanation of the observed disk structure in the HD 142527
system, especially the wide gap within 130 AU, 
but must be confirmed by continued orbital monitoring
of the binary system to confirm that it is indeed in an eccentric
orbit.  Followup SAM observations are thus absolutely critical for this system
and are still possible within the lifespan of NACO.
Eventually, a dynamical mass can be determined for this system, 
perhaps in 10 year timescales and monitoring on 1-2 year timescales
may help confirm the eccentricity of the orbit 
\citep[see~e.g.][for~a~similar~example]{Bil10}.

\section{Conclusions}

We detect a likely close companion to HD 142527 with separation 
of  88$\pm$5 mas (12.8$\pm$1.5~AU at 145 pc)
and flux ratios in H, K, and L' of   
0.016$\pm$0.007, 0.012$\pm$0.008, 0.0086$\pm$0.0011 
respectively (3-$\sigma$ errors) relative to the primary 
star and inner disk.  The companion is consistent with
mass estimates of 0.1-0.4 M$_{\odot}$ from the models of 
\citet[][]{Bar98}.  However, continued orbital monitoring will 
be necessary to provide more accurate mass estimates, 
as model masses contain significant uncertainties at these 
young ages.  The inner binary likely explains the 20 AU offset observed
by \citet[][]{Fuk06} between the primary and the disk center
\citep[][]{Pic08, Nel03}.  Additionally, the large cavity observed in the HD 142527 may be the signature 
of an unseen planet interacting with the inner binary.

\clearpage

\begin{deluxetable}{lcc}
\tabletypesize{\footnotesize}
%\rotate
\tablecaption{Properties of the HD 142527 AB System\label{tab:photom}}
\tablewidth{0pt}
\tablehead{
\colhead{} & \colhead{Primary + Disk} & \colhead{Secondary}}

\startdata

%\cutinhead{System Properties}
%Coordinates (J2000) & 18 53 05.8743 & -50 10 49.880 \\
Distance      & \multicolumn{2}{c}{145$\pm$15 pc\tablenotemark{a}} \\
Age           & \multicolumn{2}{c}{5$^{+8}_{-3}$ Myr\tablenotemark{a}} \\
Proper Motion ($\mu_{\alpha}$, $\mu_{\delta}$) & \multicolumn{2}{c}{(-11.19$\pm$0.93, $-$24.46$\pm$0.79) mas/yr\tablenotemark{b}}  \\

%\cutinhead{Astrometry}
%\cutinhead{Epoch 1, 11 Apr 2009 (UT)} 
Separation: 11 March 2012 UT           & \multicolumn{2}{c}{88$\pm$5 mas (12.8$\pm$1.5~AU)} \\ 
Position Angle: 11 March 2012 UT       & \multicolumn{2}{c}{133.3$\pm$2.5$^{\circ}$} \\
%\cutinhead{Epoch 2, 9 May 2010 (UT)} 

%\cutinhead{Photometry and Mass Estimates}
%Value & Primary & Secondary \\ \hline
Flux Ratio in $H$      & \nodata  & 0.016$\pm$0.007 \\
Flux Ratio in $Ks$ & \nodata & 0.012$\pm$0.008 \\
Flux Ratio in $L'$ & \nodata & 0.0086$\pm$0.0011 \\
%$J$ (mag)                        & 6.503$\pm$0.029\tablenotemark{c} &$\pm$ \\
%$H$ (mag)                        & 5.715$\pm$0.031\tablenotemark{c} & $\pm$ \\
%$Ks$ (mag)                       & 4.980$\pm$0.020\tablenotemark{c} & $\pm$ \\
%$L'$ (mag)                       & 3.89$\pm$\tablenotemark{c} &  $\pm$ \\
$H$ (mag)                        & 5.94\tablenotemark{c} & 10.5$\pm$0.2 \\
$Ks$ (mag)                       & 5.20\tablenotemark{c} & 10.0$\pm$0.3 \\
$L'$ (mag)                       &  3.89 \tablenotemark{c} & 9.1 $\pm$0.1 \\
$H-Ks$ (mag)                    & 0.74  & 0.5$\pm$0.4 \\
$Ks-L'$ (mag)                    & 1.31  & 0.9$\pm$0.3 \\
$M_H$ (mag)                      & 0.2$\pm$0.2 & 4.8$\pm$0.3 \\ 
$M_{K_s}$ (mag)                   & -0.6$\pm$0.2 & 4.2$\pm$0.3 \\ 
$M_{L'}$ (mag)                   & -1.9$\pm$0.2 & 3.3$\pm$0.2 \\ 
Spectral type                    &  F6IIIe   &  \nodata \\
Estimated Mass   & 2.2$\pm$0.3 M$_{\odot}$\tablenotemark{a} & 0.1-0.4 M$_{\odot}$ \\
%Estimated $T_{eff}$  & 6250\tablenotemark{a} & 2702 \\
%Estimated log(g) (from L$_{bol}$)     &  \nodata & 4.20$\pm$0.11 dex \\
\enddata
\tablenotetext{a}{\citet[][]{Ver11}}
\tablenotetext{b}{\citet[][]{vLe07}}
\tablenotetext{c}{\citet[][]{Mal98}}
%\tablenotetext{d}{Verhoeff et al. 2011}
%\tablenotetext{e}{\citet{Dan94}}

%\tablenotecomments{\textcolor{red}{**NEED TO ADD log(g) estimate**}}
\end{deluxetable}

\begin{deluxetable}{lcccc}
\tabletypesize{\footnotesize}
\tablecaption{HD 142527 2012-03-11 (UT) observation log \label{tab:obslog}}
\tablewidth{0pt}
\tablehead{
\colhead{Target} & \colhead{UT time} & \colhead{Band} & \colhead{DIT (ms)} & \colhead{NDIT} }
\startdata
HD142527 & 2012-03-11T06:02:25.1402 & L' & 120 & 330 \\ 
HD142695 & 2012-03-11T06:37:33.3990 & L' & 120 & 330  \\ 
HD142527 & 2012-03-11T06:46:55.5591 & L' & 120 & 330  \\ 
HD142384 & 2012-03-11T06:57:19.0879 & L' & 120 & 330  \\ 
HD142527 & 2012-03-11T07:07:34.3389 & L' & 120 & 330  \\ 
HD144350 & 2012-03-11T07:18:46.5699 & L' & 120 & 330  \\ 
HD142527 & 2012-03-11T07:32:05.3348 & L' & 120 & 330  \\ 
HD142695 & 2012-03-11T07:41:37.5524 & L' & 120 & 330  \\ 
HD142527 & 2012-03-11T07:51:14.5767 & L' & 120 & 330  \\ 
\hline
HD142527 & 2012-03-11T08:13:19.5884 & Ks & 109 & 360  \\ 
HD142384 & 2012-03-11T08:27:00.5782 & Ks & 109 & 360  \\ 
HD142527 & 2012-03-11T08:42:30.3669 & Ks & 109 & 360  \\ 
HD142695 & 2012-03-11T08:53:16.9986 & Ks & 109 & 360  \\ 
HD142527 & 2012-03-11T09:03:59.4995 & Ks & 109 & 360  \\ 
\hline
HD142527 & 2012-03-11T09:23:19.1784 & H  & 100 & 360  \\ 
HD142384 & 2012-03-11T09:34:21.3448 & H  & 100 & 360  \\ 
HD142527 & 2012-03-11T09:44:44.1641 & H  & 100 & 360  \\ 
\enddata 
\end{deluxetable}

\begin{figure}
\includegraphics[width=2in]{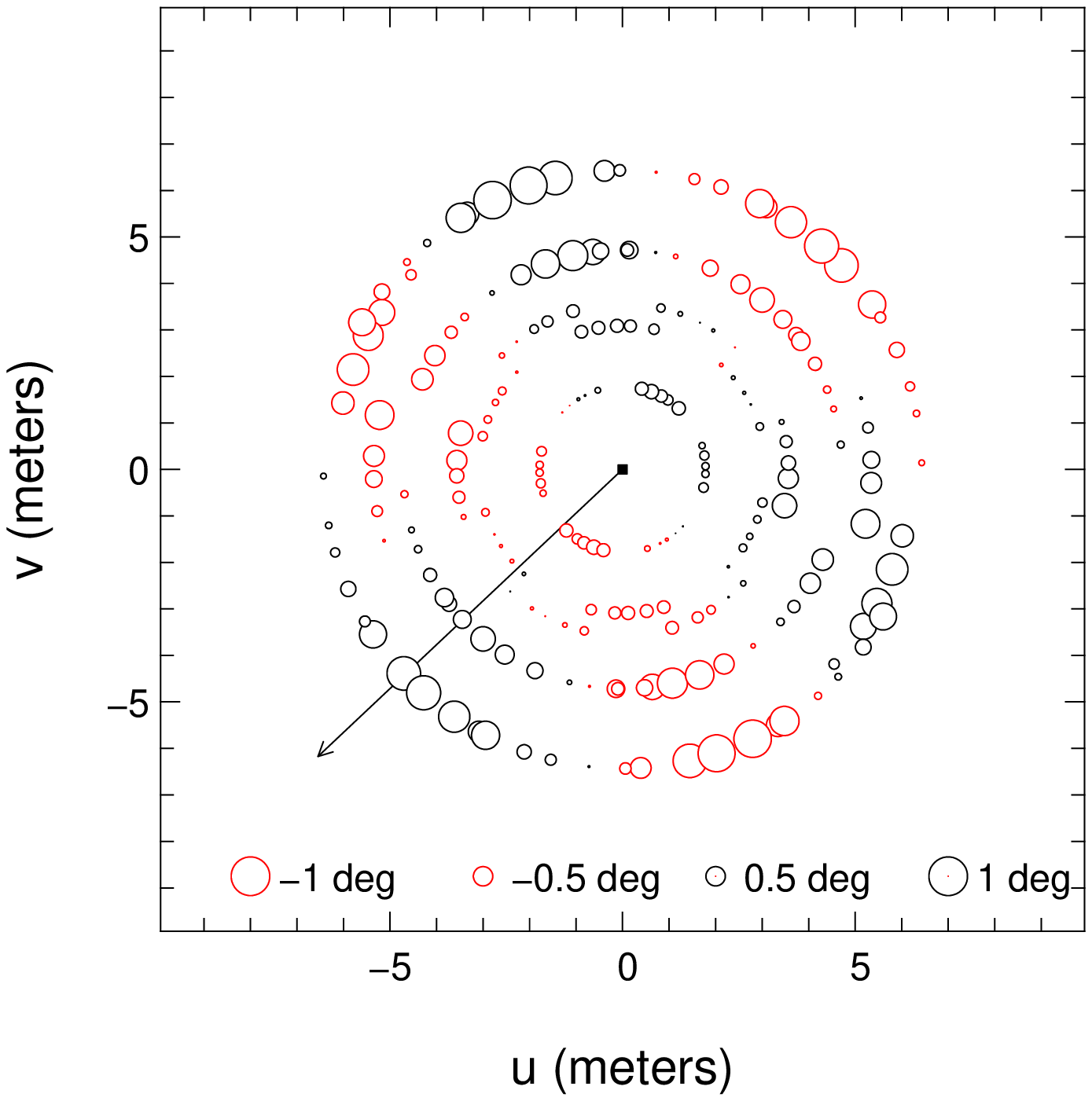}
\includegraphics[width=2.2in]{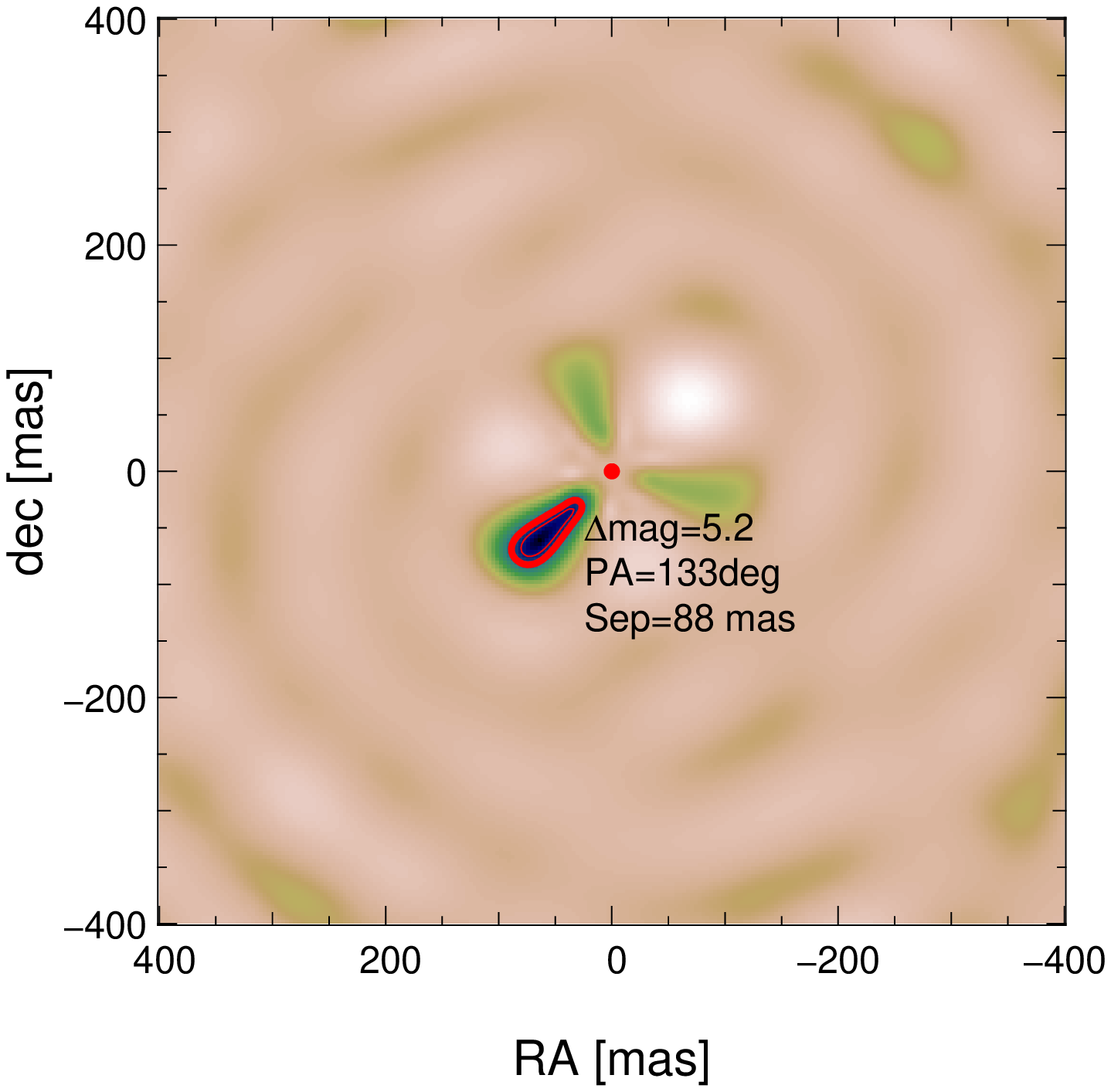} 
\includegraphics[width=2.5in]{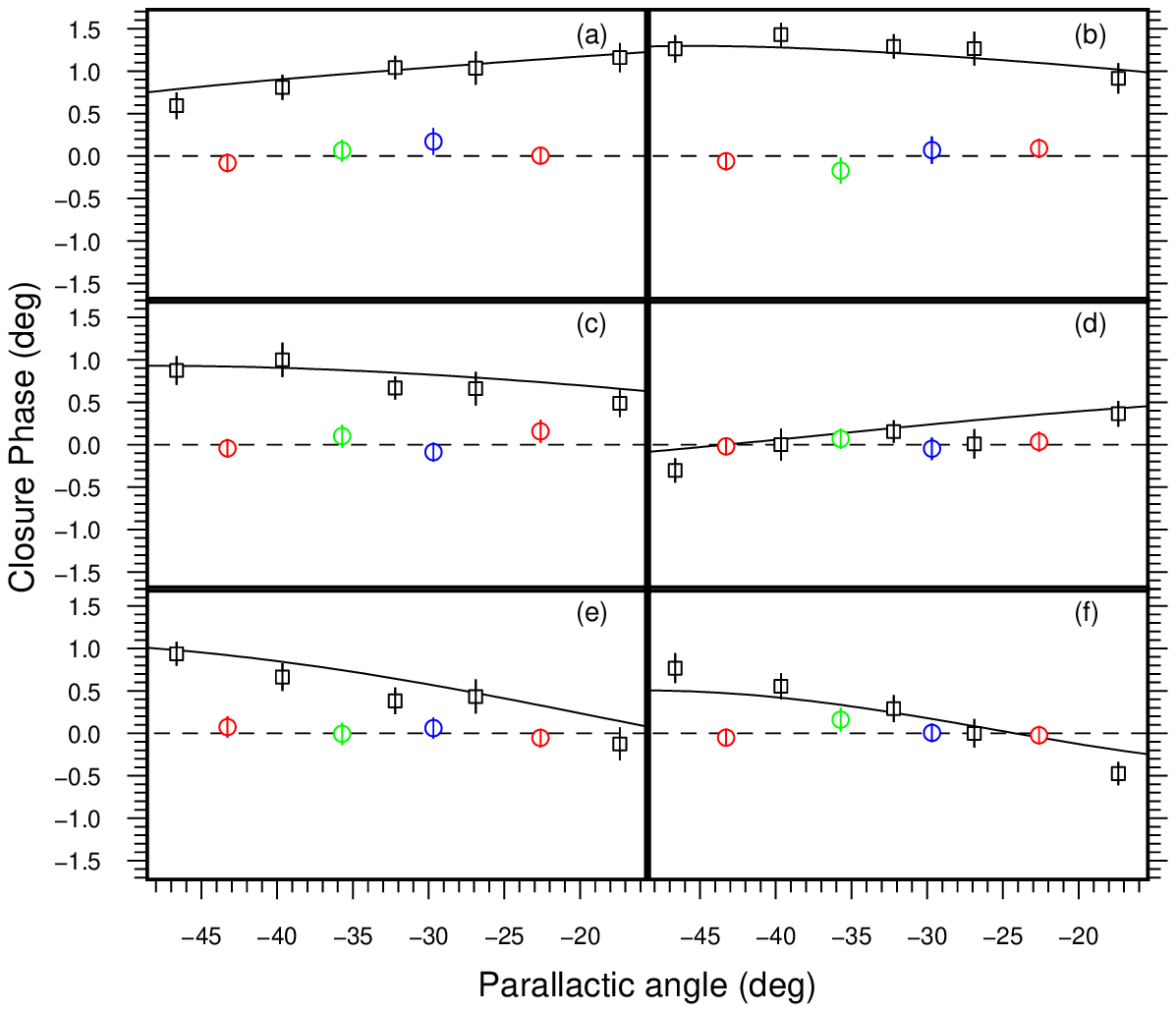} 
\caption{Left panel: L' band UV coverage on HD142527. The size and colors of the
markers are relative to the phase measured. The larger the size, the
higher the value of the phase. The colors denotes the sign of the phase
(red are negatives values).  The plot shows diagonal stripes orthogonal to the direction of the
binary companion (indicated by the arrow). 
Middle panel: L' band $\chi^2$ surface as a function of RA and DEC obtained from
the best fit binary model to the closure phases. A clear minimum
indicates the position of the stellar companion, coherent with the
orientation of the stripes in the Fourier domain. The red contours
correspond to 3 and 5 sigma error bars in the detection.
Right panel: L' band closure phase as a function of parallactic angle for the 6 largest
3-hole triangles.  Calibrator data are represented as colored squares (red HD142695, green
HD144350 and blue HD142384), while HD142527 data are plotted as black
squares.  The solid line is the closure phase predicted by the best fitting binary
system model. 
\label{fig:image}
}
\end{figure}

%\clearpage

%\begin{figure}
%\includegraphics[width=6in]{fitHD142.eps} 
%\caption{Closure phase as a function of parallactic angle for the 16 largest
%3-hole triangles.  Calibrator data are represented as colored squares (red HD142695, green
%HD144350 and blue HD142384), while HD142527 data are plotted as black
%squares. The dashed line is the polynomial fit of the calibrator, and
%the solid line is the closure phase predicted by the best fitting binary
%system model. The same fit quality is obtained on all 35 closure phases.
%\label{fig:closure}
%}
%\end{figure}

%\clearpage

\begin{figure}
\includegraphics[width=6in]{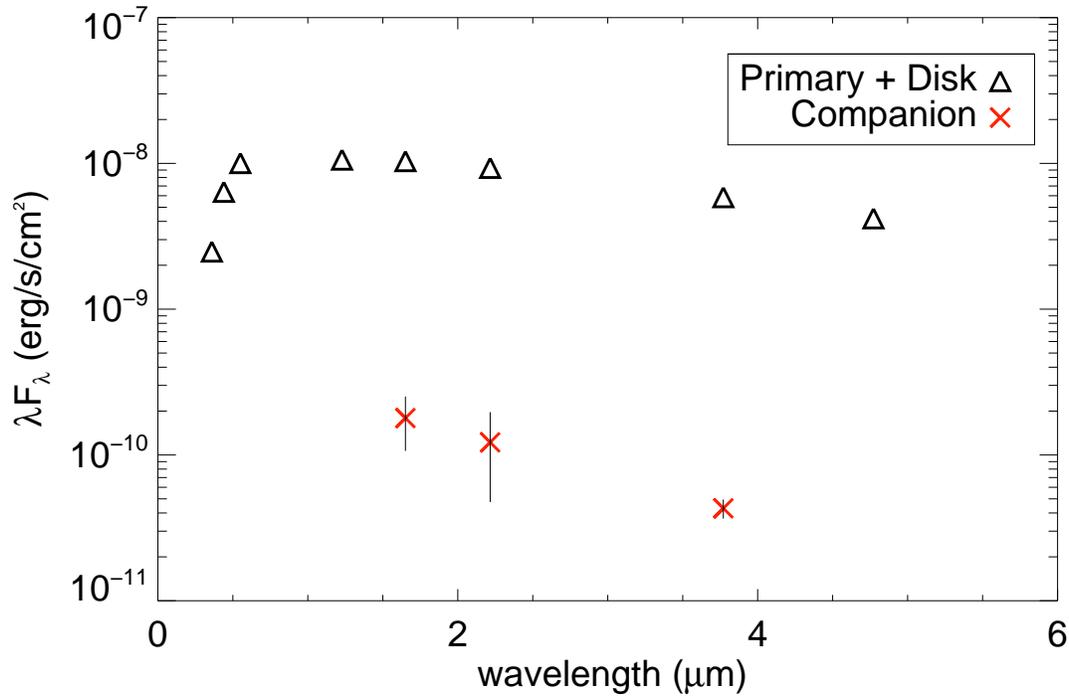}
\caption{SED for HD 142527 as well as companion fluxes
in the same bandpasses.  SED datapoints are drawn 
from the photometry of \citet[][]{Mal98}.
\label{fig:sed}
}
\end{figure}

%\clearpage

\begin{figure}
\includegraphics[width=3in]{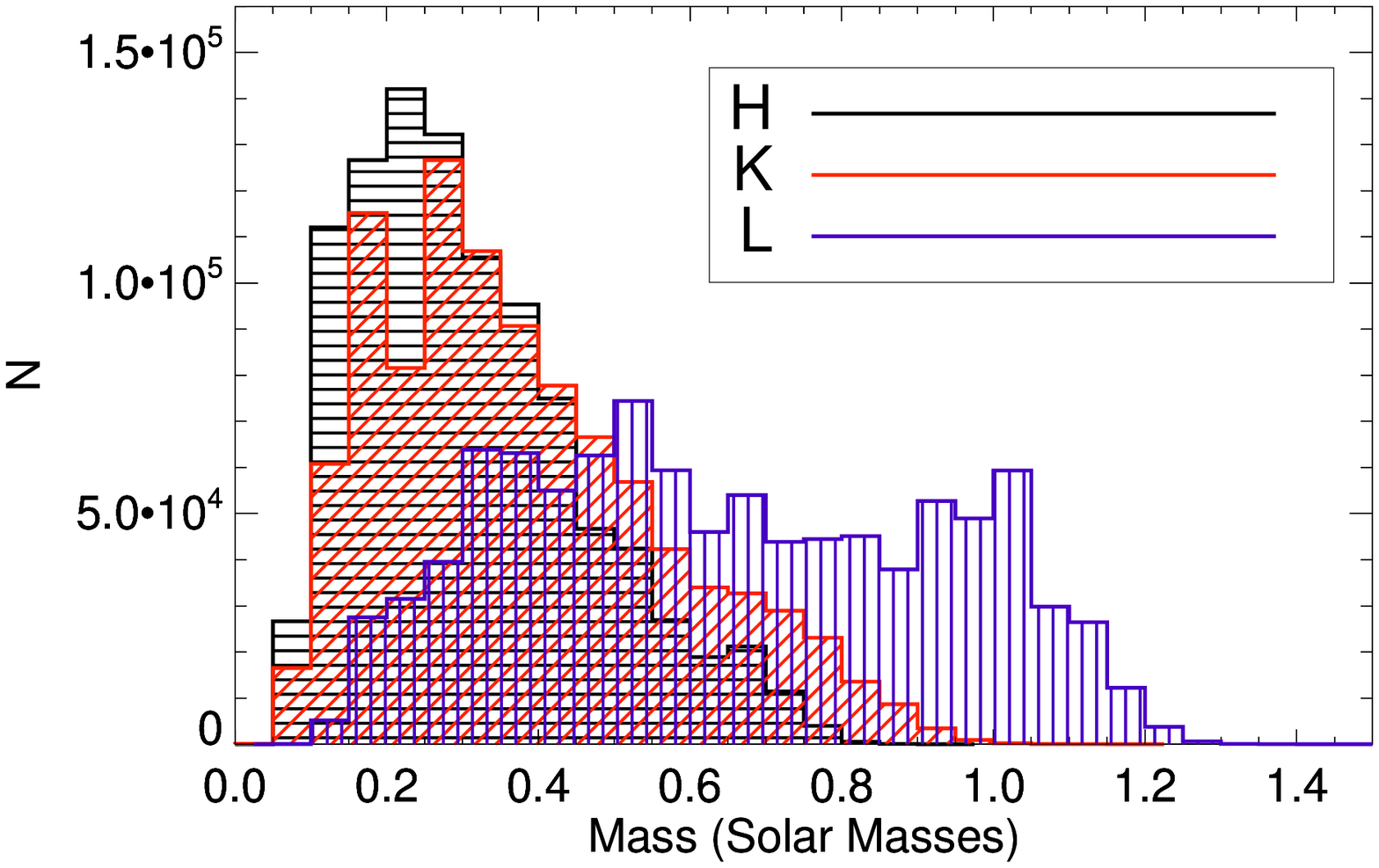}
\includegraphics[width=3in]{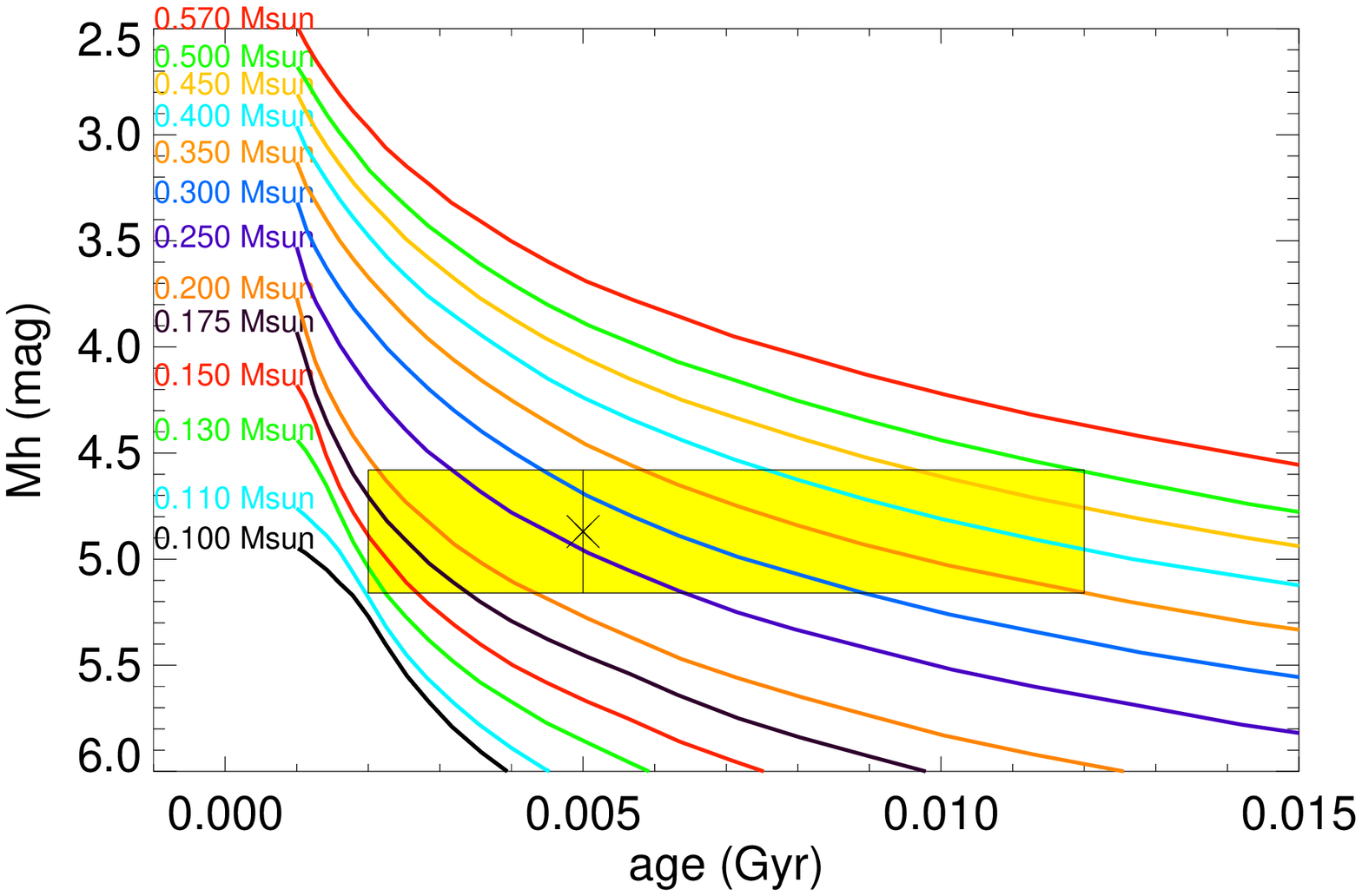}
\caption{Left: Mass estimate histograms for HD 142527B.
We adopt Monte Carlo methods to account for the range
of possible ages for this system.  An ensemble of 10$^{6}$ possible ages are drawn
from a Gaussian in log space, centered on log(age) = 6.7 and with 
$\sigma$(log(age)) = 0.4.  We then interpolate with age and single
band absolute magnitude to find the best mass for the companion 
from the models of \citet[][]{Bar98}.  We plot here 
the resulting mass distributions from single-band 
absolute magnitudes in H, K, and L'. 
It is instantly apparent that the mass estimate for the companion is not 
well constrained at these ages but is most likely to lie in the range from
0.1-0.4 M$_{\odot}$.  Right: Age vs. absolute magnitude in the H band.  
We plot isomass contours \citep[from~the~models~of][]{Bar98}
as a function of age and absolute magnitude, as well as the 
assumed age and absolute magnitude of the companion.
The models themselves are highly dependent on age; the 
region constrained by the observations (yellow rectangle)
is consistent with companion masses from 0.1 - 0.4 M$_{\odot}$.
\label{fig:masshist}
}
\end{figure}


\begin{thebibliography}{}
\bibitem[Acke $\&$ van den Ancker(2004)]{Ack04} Acke, B., \& van den Ancker, M.~E.\ 2004, \aap, 426, 151 
\bibitem[Andrews et al.(2011)]{And11} Andrews, S.~M., Wilner, D.~J., Espaillat, C., et al.\ 2011, \apj, 732, 42 
\bibitem[Baines et al.(2006)]{Bai06} Baines, D., Oudmaijer, R.~D., Porter, J.~M., \& Pozzo, M.\ 2006, \mnras, 367, 737 
\bibitem[Baraffe et al.(1998)]{Bar98} Baraffe, I., Chabrier, G., Allard, F., \& Hauschildt, P.~H.\ 1998, \aap, 337, 403 
\bibitem[Biller et al.(2010)]{Bil10} Biller, B.~A., Liu, M.~C., Wahhaj, Z., et al.\ 2010, \apjl, 720, L82 
\bibitem[Brandner et al.(2000)]{Bra00} Brandner, W., Zinnecker, H., Alcal{\'a}, J.~M., et al.\ 2000, \aj, 120, 950 
\bibitem[Brown et al.(2009)]{Bro09} Brown, J.~M., Blake, G.~A., Qi, C., et al.\ 2009, \apj, 704, 496 
\bibitem[de Zeeuw et al.(1999)]{deZ99} de Zeeuw, P.~T., Hoogerwerf, R., de Bruijne, J.~H.~J., Brown, A.~G.~A., 
\& Blaauw, A.\ 1999, \aj, 117, 354 
\bibitem[Dupuy et al.(2010)]{Dup10} Dupuy, T.~J., Liu, M.~C., Bowler, B.~P., et al.\ 2010, \apj, 721, 1725 
\bibitem[Elias et al.(1982)]{Eli82} Elias, J.~H., Frogel, 
J.~A., Matthews, K., \& Neugebauer, G.\ 1982, \aj, 87, 1029 
\bibitem[Fukagawa et al.(2006)]{Fuk06} Fukagawa, M., Tamura, M., Itoh, Y., et al.\ 2006, \apjl, 636, L153 
\bibitem[Henize (1976)]{Hen76} Henize, K.~G.\ 1976, \apjs, 30, 491 
\bibitem[Houk (1978)]{Hou78} Houk, N.\ 1978, Ann Arbor : Dept.~of Astronomy, University of Michigan : distributed by University 
Microfilms International, 1978-,  
\bibitem[Hu\'elamo et al.(2011)]{Hue11} Hu{\'e}lamo, N., Lacour, S., Tuthill, P., et al.\ 2011, \aap, 528, L7 
\bibitem[Kley et al.(2008)]{Kle08a} Kley, W., Papaloizou, J.~C.~B., $\&$ Ogilvie, G.~I.\ 2008, \aap, 487, 671 
\bibitem[Kley $\&$ Nelson(2008)]{Kle08b} Kley, W., \& Nelson, R.~P.\ 2008, \aap, 486, 617 
\bibitem[Kley $\&$ Nelson(2012)]{Kle12} Kley, W., \& Nelson, R.~P.\ 2012, arXiv:1203.1184, to appear in ARAA. 
\bibitem[Kraus $\&$ Ireland(2012)]{Kra12} Kraus, A.~L., \& Ireland, M.~J.\ 2012, \apj, 745, 5 
\bibitem[Kurucz(1991)]{Kur91} Kurucz, R.~L.\ 1991, Precision Photometry:  Astrophysics of the Galaxy, 27 
\bibitem[Lacour et al.(2011a)]{Lac11a} Lacour, S., Tuthill, P., 
Ireland, M., Amico, P., \& Girard, J.\ 2011, The Messenger, 146, 18 
\bibitem[Lacour et al.(2011b)]{Lac11b} Lacour, S., Tuthill, P., Amico, P., et al.\ 2011, \aap, 532, A72 
%\bibitem[Larwood $\&$ Kalas(2001)]{Lar01} Larwood, J.~D., \& Kalas, P.~G.\ 2001, \mnras, 323, 402 
\bibitem[Malfait et al.(1998)]{Mal98} Malfait, K., Bogaert, E., \& Waelkens, C.\ 1998, \aap, 331, 211 
\bibitem[Najita et al.(2007)]{Naj07} Najita, J.~R., Strom, S.~E., \& Muzerolle, J.\ 2007, \mnras, 378, 369 
\bibitem[Nelson(2003)]{Nel03} Nelson, R.~P.\ 2003, \mnras, 345, 233 
\bibitem[Pichardo et al.(2008)]{Pic08} Pichardo, B., Sparke, L.~S., \& Aguilar, L.~A.\ 2008, \mnras, 391, 815 
\bibitem[Pott et al.(2010)]{Pot10} Pott, J.-U., Perrin, M.~D., Furlan, E., et al.\ 2010, \apj, 710, 265 
%\bibitem[Regaly et al.(2012)]{Reg12} 
\bibitem[Robin et al.(2003)]{Rob03}  Robin, A.~C., Reyl{\'e}, C., Derri{\`e}re, S., \& Picaud, S.\ 2003, \aap, 409, 523 
\bibitem[Siess et al.(2000)]{Sie00} Siess, L., Dufour, E., \& Forestini, M.\ 2000, \aap, 358, 593 
\bibitem[Teixeira et al.(2000)]{Tei00} Teixeira, R., Ducourant, C., Sartori, M.~J., et al.\ 2000, \aap, 361, 1143 
\bibitem[Tokunaga(2000)]{Tok00} Tokunaga, A. 2000, in Allen's Astrophysical Quantities, pg. 152, edited by A.N. Cox
\bibitem[Tuthill et al.(2000)]{Tut00} Tuthill, P.~G., Monnier, J.~D., \& Danchi, W.~C.\ 2000, \procspie, 4006, 491 
\bibitem[Tuthill et al.(2006)]{Tut06} Tuthill, P., Lloyd, J., Ireland, M., et al.\ 2006, \procspie, 6272,  
\bibitem[Tuthill et al.(2010)]{Tut10} Tuthill, P., Lacour, S., Amico, P., et al.\ 2010, \procspie, 7735,  
%\bibitem[van den Ancker et al.(1998)]{vdA98} van den Ancker, M.~E., de Winter, D., \& Tjin A Djie, H.~R.~E.\ 1998, \aap, 330, 145 
\bibitem[van Boekel et al.(2004)]{Boe04} van Boekel, R., Min, M., Leinert, C., et al.\ 2004, \nat, 432, 479 
\bibitem[van Leeuwen(2007)]{vLe07} van Leeuwen, F.\ 2007, \aap, 474, 653 
\bibitem[Verhoeff et al.(2011)]{Ver11} Verhoeff, A.~P., Min, M., Pantin, E., et al.\ 2011, \aap, 528, A91 
\bibitem[Waelkens et al.(1996)]{Wae96} Waelkens, C., Waters, L.~B.~F.~M., de Graauw, M.~S., et al.\ 1996, \aap, 315, L245 
\end{thebibliography}
\end{document}